\documentstyle[12pt,eqsecnum,aps,prd,preprint]{revtex}
\input epsf.tex
\begin{document}
\preprint{\baselineskip=18pt \vbox{\hbox{SU-4240-654}\hbox{January 1997}}}
\title{\large\bf Multiflavor Massive Schwinger Model With Non-Abelian
 Bosonization}
\author{David Delphenich and Joseph Schechter}
\address{Physics Department, Syracuse University\\ 
Syracuse, NY 13244-1130} 
\maketitle
\begin{abstract}
We revisit the treatment of the multiflavor massive Schwinger model by
non-Abelian Bosonization.  We compare three different approximations to the
low-lying spectrum:  i) reading it off from the bosonized Lagrangian
(neglecting interactions), ii) semi-classical quantization of the static
soliton, iii) approximate semi-classical quantization of the ``breather''
solitons.  A number of new points are made in this process.  We also suggest
a different ``effective low-energy Lagrangian'' for the theory which permits
easy calculation of the low-energy scattering amplitudes.  It correlates
an exact mass formula of the system with the requirement of the Mermin-Wagner
theorem.
\end{abstract}
\vskip 1cm
\section {Introduction}
 The study of two-dimensional field theories has been extremely useful
 \cite{QED2} for  understanding many aspects of the realistic four-dimensional
 cases.  In a very interesting paper \cite{Co}, Coleman analyzed the 
multi-flavor generalization of two-dimensional electrodynamics\cite{Sch}.  The well-known Lagrangian density is
\begin {equation}
{\cal L}=-\frac{1}{4}F_{\mu\nu}F_{\mu\nu}-{\bar\psi}_f[\gamma_\mu(\partial_\mu-ie{\cal A}_\mu)+m^\prime]\psi_f,
\end {equation}
where $F_{\mu\nu}=\partial_\mu{\cal A}_\nu-\partial_\nu{\cal A}_\mu$ and
 summation is to be understood over the flavor index f. (Here
 ${\bar\psi}_f=\psi^\dag_f\gamma_2$ and one may choose $\gamma_1=\sigma_1,
 \gamma_2=\sigma_2$.) Since the electric charge has the dimension of mass in
 this theory it is meaningful to define the strong coupling regime as
\begin{equation}
\ e >> m^\prime,
\end{equation} 
where $m^\prime$ is the common fermion mass. (It is also interesting
to allow different masses, $m_f$ for each fermion.)

The natural presentation \cite{Co} of the theory in the strong coupling
regime is its bosonized form \cite{Co75}.  Then the large quantity $e^2$  ends up just multiplying a quadratic (mass) term and does not complicate the interactions.  The resulting Lagrangian has a lot of similarity to the low energy effective meson Lagran
gian used for describing QCD.  Since some exact results are known for the two-dimensional case we may hope to learn more about various aspects of the QCD effective Lagrangian.  That is, in fact, our motivation for looking at this model and sets the framew
ork of our analysis. 

Coleman \cite{Co} used an Abelian bosonization technique and
showed that the lowest state in the 2-flavor model is a "meson" with quantum
numbers  $I^{PG}= 1^{-+}$ \footnote{$G = e^{i\pi I_y}C$, where C is the
charge parity, is the usual G parity.  Note that  ${\bar\psi}\gamma_5\psi$
goes to $-{\bar\psi}\gamma_5\psi$ under charge conjugation
($\gamma_5=-i\gamma_1\gamma_2$ here), unlike the four-dimensional case.}.  He
pointed out that the first excited state has the quantum numbers  $I^{PG}=
0^{++}$ and obeys the exact mass relation 
\begin{equation}
m(0^{++})=\sqrt{3} m(1^{-+}).
\end{equation}
In addition, there are an infinite number of unstable mesons in the model.
At a much larger mass scale there appears the  $I^{PG}= 0^{--}$ meson, which
would lie rather low in the weak coupling limit \cite{Co}.

A complicating feature in the treatment of \cite{Co} is that the lowest-lying
physical states emerge in a very asymmetrical manner.  The members of this
$I^{PG}=1^{-+}$ triplet, in fact, variously emerge as a soliton, an
anti-soliton and a soliton-anti-soliton bound state (or "breather").  It is
possible to give a symmetrical treatment by using the more recently
discovered non-Abelian bosonization technique \cite{Wi}.  Gepner \cite{Ge}
carried out this analysis, showing that the $1^{-+}$ triplet could be treated
symmetrically as the collective excitation of the classical soliton solution
in the non-Abelian model.  This method of treating the meson states is
similar to that employed in the treatment \cite{Gua} of three-flavor baryons in the four-dimensional Skyrme model \cite{Sky}. 

In the present note we shall investigate some aspects of the non-Abelian bosonization of the model in more detail.  As a preliminary, we point out that some interesting things can be said about the low-lying $1^{-+}$ triplet at the level of the non-Abelia
n Lagrangian itself, without going to the soliton sectors.  In this way, for example, we may easily relate two of Coleman's "three things I don't understand" \cite{Co} to the situation in the QCD meson spectrum.  We address a problem concerning the true l
owest-lying state which appeared in \cite{Ge}.  There it was found that, at the semi-classical level, $m(0^{++}) < m(1^{-+})$, which would make the $I^{PG}=0^{++}$ meson lowest-lying.  This was interpreted as a deficiency of the approximation in treating 
the breather modes.  We investigate further the breather modes here and develop a quantitative approximation method for treating their excitations.  We point out that a natural alternative interpretation of the model yields $m(0^{++})>m(1^{-+})$, in agree
ment with Coleman.  This is welcome since the semi-classical treatment of soliton collective modes has usually given a nice understanding of at least the overall features of the baryon spectrum.  A procedural difference from \cite{Ge} here, which yields t
he same result, involves starting from the free bosonized theory and then gauging it, rather than bosonizing the interacting theory as a whole.  We also give a slightly different treatment of the soliton collective quantization.

Finally, we investigate the possibility of an approximate low-energy
effective Lagrangian description of multiflavor $QED_2$ rather than the
exact bosonized description.  A low-energy effective Lagrangian has an
advantage over the exact bosonized theory in that it can contain all the
low-lying particles.  Hence the tree-level scattering amplitudes computed
from this Lagrangian should be good approximations at low energy.  Furthermore
we show that taking the linear sigma model as the effective Lagrangian leads
to a correlation between the special mass formula (1.3) and the Mermin-Wagner
theorem \cite{MW,Shif} on the impossibility of the spontaneous breakdown of
a continuous symmetry in two dimensions.

In section 2 we show how the non-Abelian bosonized multiflavor $QED_2$
Lagrangian can be derived by a suitable ``gauging'' of Witten's bosonized
Lagrangian \cite{Wi} representing a multiplet of free fermi fields.  Section
3 contains a discussion showing how certain puzzling features of the
multiflavor theory can be understood at the tree level of the resulting
theory.  The analogy to low-energy particle physics phenomena is pointed
out.  We go beyond the tree approximation by exploiting the semi-classical
quantization of the classical solitons of the model.  The well-known
time-independent and time-dependent (breather) solitons are discussed in
section 4.  Section 5 contains a treatment of the semi-classical quantization
of the static solitons.  In section 6 the same method is applied to the
time-dependent solitons by making a kind of Born-Oppenheimer approximation
which requires computing the time-averaged ``moment of inertia'' of the
soliton.  Details of this calculation are given in Appendix A.  Section 7
contains a comparison of the alternative approaches to the spectrum given in
sections 3, 5, and 6.  The need for an effective Lagrangian is explained and
it is argued that the linear sigma model is a suitable candidate.  It is
shown to lead to an understanding of the mass relation (1.3) and is used to
find the low-energy scattering amplitude.

\section{Bosonized action}
First, we shall write down the bosonized version of the free fermion terms in
the Lagrangian (2.1).  It is built from the $N_f\times N_f$ unitary matrix field U(x) which transforms as
\begin{equation}
U(x) \to U_L U(x) U^{-1}_R
\end{equation}
under the global chiral $U_L(N_f) \times U_R(N_f)$
transformation $\psi_{L,R}\to U_{L,R} \psi_{L,R}$.  Here
$\psi_{L,R}=\frac{1}{2}(1\pm\gamma_5)\psi$.  There are three pieces:
\begin{equation}
\Gamma_{free} = \Gamma_\sigma + \Gamma_m + \Gamma_{WZW}.
\end{equation} 
$\Gamma_\sigma$ and $\Gamma_m$ are essentially the usual kinetic and mass
terms of the non-linear sigma model:
\begin{equation} 
\Gamma_\sigma+\Gamma_m= \int d^2 x [-\frac{1}{8\pi}
Tr(\partial_\mu U\partial_\mu U^\dag)+
\frac{1}{2} m^2 Tr(U+U^\dag-2)],
\end{equation} 
where \cite{QED2} $m$ is essentially proportional to $m^\prime$ in (1.1).
Clearly the first term in (2.3) is chiral invariant while the second has the
same chiral transformation property as the mass term in (1.1).  The
characteristic Wess-Zumino-Witten term necessary for non-Abelian bosonization
\cite{Wi} may be compactly written, using the matrix one-form $\alpha= dU U^\dag$, as
\begin{equation}
\Gamma_{WZW} = \frac{1}{12\pi}\int_{M^3} Tr(\alpha^3),
\end{equation}
where $M^3$ is a three-dimensional manifold whose boundary is the
two-dimensional Minkowski space. 
 
Now let us "gauge" the set of $N_f$ bosonized massive Dirac fields
represented by (2.2).  We can always include a gauge-invariant piece
$\Gamma_\gamma$ containing just the electromagnetic fields:
\begin{equation} 
\Gamma_\gamma=\int d^2 x \left\{ -\frac{1}{4} F_{\mu\nu}
F_{\mu\nu}+\frac{ie\theta}{4\pi}\epsilon_{\mu\nu} F_{\mu\nu}\right\}.
\end{equation}
The second term, labeled by the angular parameter $\theta$, describes the
effect of a background electric field \cite{Co}.  It violates parity
invariance and is the analog of the $\theta$ parameter in 4-dimensional QCD
\cite{Thooft}.  We shall, for the most part, consider only the $\theta=0$ case
in the present paper. Finally, and most importantly, we must include the
matter-gauge field interaction.  At the fermion level it is, of course,
obtained by replacing $\partial_\mu \psi_f$ by $(\partial_\mu - ie {\cal
A}_\mu)\psi_f$ so that the change in $\partial_\mu \psi_f$ under a local U(1)
gauge transformation $\psi_f\rightarrow e^{i\Lambda(x)}\psi_f$ is canceled by
the transformation of the gauge field ${\cal A}_\mu\rightarrow {\cal A}_\mu +
\frac{1}{e}\partial_\mu\Lambda$.  At the bosonic level there is a problem
with this approach since the basic field $U(x)$ represents only electrically
neutral objects (mesons) and should thus remain invariant under a U(1) gauge
transformation.  It would, at first, seem that the free bosonized action
(2.2) is gauge-invariant as it stands so there is no need to couple it to the
U(1) gauge field ${\cal A}_\mu$.  We seem to have reached a dead end! 
 
However, the situation for the three-dimensional term (2.4) is not really
clear and, in any event, we have the obligation to demand a consistent
gauging of the bosonized massless free field terms
$\Gamma_\sigma+\Gamma_{WZW}$ with respect to non-Abelian flavor
transformations (under which $U$ does transform).  We shall thus add terms to
make $\Gamma_{WZW}$ invariant under local $U(N_f)$ vector-type
transformations and afterwards specialize to the electromagnetic
$U(1)_{EM}$ subgroup.  Under a local infinitesimal
$U(N_f)$ vector-type transformation one has
\begin{equation} 
\delta U = i[E, U],\quad   \delta A = \frac{1}{e} dE + i[E, A],
\end{equation} 
where $E=E^\dag$ and A is the matrix one-form of $U(N_f)$ gauge
fields.  Eq. (2.4) can now be gauged iteratively \cite{KRS}.  Its variation
under (2.6) is seen, with the help of Stokes's theorem to be partially
canceled by the variation of the additional term $\frac{ie}{4\pi} \int_{M^2}
Tr[A(\alpha+\beta)]$, where $\beta=U^\dag dU$.  The remaining variation of the
new term is canceled by the variation of the term $\frac {e^2}{4\pi}
\int_{M^2} Tr[UAU^\dag A]$  and the procedure terminates.  Now if we
specialize $A_\mu$ to the desired U(1) component by setting $A_\mu = {\cal A}_\mu {\bf 1}$ we see that the second new term vanishes while the first becomes the interaction term
\begin{equation}
\Gamma_{int}= - \frac{e}{2\pi} \int d^2 x\ \epsilon_{\alpha\beta}
{\cal A}_\alpha Tr(\partial_\beta UU^\dag).
\end{equation} 
The total bosonized action for multiflavor QED is then the sum of (2.3), (2.4), (2.5) and (2.7).  As a check on this procedure we may calculate the electromagnetic current 
\begin{equation}
J^{EM}_\mu = \frac{\delta\Gamma}{\delta {\cal A}_\mu}
|_{{\cal A}_\mu=0}=-\frac{e}{2\pi}\epsilon_{\mu\nu} Tr(\partial_\nu U
U^\dag).
\end{equation} 
The action may be further simplified by making use of the fact that there is
no propagating photon degree of freedom in the two-dimensional theory;  then
the photon field may be "integrated out."  This is conveniently accomplished
by the substitution \cite{EFHK} $F_{\mu\nu}=\epsilon_{\mu\nu}F$.  The field F obeys the equation of motion:
\begin{equation} 
F = \frac{ie}{2\pi}(\theta+ i\ ln\ det\ U),
\end{equation} 
wherein $Tr(\partial_\nu UU^\dag) = \partial_\nu ln\ det\ U$ was used.
Substituting (2.9) back into $\Gamma$ gives
\begin{equation} 
\Gamma = \int d^2 x [-\frac{1}{8\pi} Tr(\partial_\mu
U\partial_\mu U^\dag)+\frac{m^2}{2} Tr(U+
U^\dag-2)-\frac{e^2}{8\pi^2}(\theta+ i\ ln\ det\
U)^2]+\Gamma_{WZW}.
\end{equation}

\section{Analogy to particle physics}
The form (2.10) can nowadays be recognized as essentially identical to that
of the four-dimensional Lagrangian describing the pseudoscalar mesons.
Coleman \cite{Co} suspected the analogy and pointed out features which were
puzzling (stated as "questions I don't understand") on the fermion picture.
However, the connection was slightly obscured by the use of the Abelian
bosonization.  Hence it may be interesting to
briefly discuss this here.  Let us simplify to the two-flavor case and set $\theta=0$.  Introduce the decomposition:
\begin{equation}
U =exp(i\sqrt{4\pi}\phi),\qquad
\phi=
\left(\begin{array}{cc}
\phi_{11}&\phi_{12}\\
\phi_{21}&\phi_{22}
\end{array}\right),
\end{equation}
with the picturesque names 
\begin{equation}
\pi^{+}=\phi_{12}\quad, \pi^{-}=\phi_{21},
\quad \pi^{0}=\frac{1}{\sqrt{2}}(\phi_{11}-\phi_{22}),
\quad \eta=\frac{1}{\sqrt{2}}(\phi_{11}+\phi_{22}).
\end{equation}
(In the present model, of course, all these fields describe electrically neutral particles.)
Then the expansion of (2.10) in powers of $\phi$ yields the quadratic terms 
\begin{equation}
\Gamma = - \int d^{2}x\left\{\partial_{\mu} \pi^{+}\partial_{\mu} \pi^{-}
+ \frac{1}{2}(\partial_{\mu}\pi^0)^{2} +
\frac{1}{2}(\partial_{\mu}\eta)^2 +
m^2_{\pi}(\pi^{+}\pi^{-}+\frac{1}{2}\pi^{0}\pi^{0})
+\frac{1}{2}m^2_{\eta}\eta^2\right\}+...,
\end{equation}
where 
\begin{equation}
m_{\pi} = 2\sqrt{\pi} m \approx 3.54m,
\quad m_{\eta} = \sqrt{4\pi m^2+\frac{2e^2}{\pi}}.
\end{equation}
In the strong coupling limit, $e>>m$, the mass of the $I^{PG} = 0^{--}$ particle $\eta$ is clearly many orders of magnitude higher than the mass of the
 $I^{PG} = 1^{-+}$ triplet.  Since $e^2$ only makes its appearance in the $\eta$ mass term, all of the other low-lying states will be of the same order of magnitude as the $\pi$ triplet (Coleman's second question). The $\eta$ essentially decouples.  In QC
D, the fact that the $\eta'$ meson is much heavier than the $\pi$ triplet is usually attributed to instanton effects rather than quark annihilation graphs \cite{DGH}. 

It is also interesting to consider what happens when we allow different masses for the fundamental fermions.  In the bosonized picture of the two-flavor model this corresponds to the additional term
\begin{equation}
\Gamma_{\Delta} = \Delta Tr[\tau_3(U+U^{\dag})],
\end{equation}
where $\Delta$ is an isospin violation parameter with dimension $(mass)^2$
and $\tau_3$ is the Pauli matrix.  Now Coleman's first question is:  why does
the $\pi$ triplet remain degenerate even if, for example,
$\sqrt{|\frac{\Delta}{m^2}|}$ has order of magnitude 10?   In the present framework it is easy to see that this is just a variant of the second question discussed above. 

We consider the strong coupling situation where $e>> \{m, \sqrt{|\Delta|}\}$.
With the decomposition (3.1), $\Gamma_{\Delta}$ expands out as
\begin{equation} 
\Gamma_\Delta = - \int d^2 x [2\sqrt{2} \Delta\eta\pi^0 +
...]\ .
\end{equation}
This mixing between the 
$\pi^0$ and $\eta$ fields requires us to diagonalize the matrix 
\begin{equation}
\left( \begin{array}{cc}	
m^2_{\pi} & 2\sqrt{2}\Delta \\
2 \sqrt{2}\Delta & m^2_{\eta}
\end{array}\right),
\end{equation}
(where $m_\pi$ and $m_\eta$ are given by (3.4)) in order to obtain the
physical $\pi^0$ and $\eta$ states and masses.  The eigenvalues of (3.7) give the physical masses
\begin{equation} 
m^2_{phys} (\pi^0)\approx m^2_\pi - \frac{\Delta^2}{m^2_\eta},
\qquad m^2_{phys}(\eta)\approx m^2_\eta + \frac{\Delta^2}{m^2_\eta},
\end{equation}
which leads to
\begin{equation}
m^2_{phys}(\pi^{\pm})-m^2_{phys}(\pi^0)\approx \frac{\Delta^2}{m^2_\eta}.
\end{equation}
Remembering that we are working in the strong coupling approximation where
$\Delta <<  m^2_\eta\approx \frac{2e^2}{\pi}$, we see that the $\pi^\pm-\pi^0$ mass splitting vanishes as $e\to \infty$!  This is essentially the same as the
effect in four-dimensional QCD that the $\pi^\pm-\pi^0$  mass splitting is due
to photon exchange diagrams rather than to the difference between the down
and up quark masses, $m_d - m_u$.  More precisely, the piece due to ($m_d -
m_u$) is proportional to $(m_d - m_u)^2$ and hence negligible (as
in (3.9) above) rather than being proportional to ($m_d-m_u$).  This also
follows from the isospin transformation properties of the quark mass
operator.  By Bose statistics, the $\pi^\pm-\pi^0$ mass difference can only be mediated by an operator satisfying $\Delta$I=2.  However the quark mass terms transform as a linear combination of $\Delta$I=0 and $\Delta$I=1 pieces. 

\section{Classical solutions}

We are interested in studying the strong coupling spectrum of the model by
quantizing the excitations around exact classical solutions \cite{Raj}..  We
adopt the ansatz \cite{EFHK} for classical solutions:
\begin{equation}
U_{c}(x,t)=diag[exp(i2\sqrt{\pi}\chi_1(x,t)),...,exp(i2\sqrt{\pi}\chi_{N_f}(x,t))].
\end{equation}
It is being assumed that $U_{c}$ depends only on $x_1$ and $x_2=it$, not
on $x_3$, the coordinate appearing in the three-dimensional term
$\Gamma_{WZW}$, eq. (2.4).  The structure of $\Gamma_{WZW}$ then shows that
it will give zero when (4.1) is substituted into it.  Hence, substituting
(4.1) into the total action (2.10) yields, after the usual Legendre
transform, the classical Hamiltonian density,
\begin{equation}
{\cal H}_{class}=\frac{1}{2}\sum^{N_f}_{i=1}(\dot
\chi^2_i+(\chi^{\prime}_i)^2)+\frac{e^2}{2\pi}(\sum_i\chi_i-\frac{\theta}
{2\sqrt{\pi}})^2+\sum_i m^2_i [1-cos(2\sqrt{\pi}\chi_i)],
\end{equation}
where we have allowed for $N_f$ different masses, $m_i$ \footnote{An
interesting discussion of the unequal mass case has been given in
\cite{EFHK}, whereas \cite{Co} and \cite{Ge} confine their attention to the
equal-mass case.}.  (Here
$\dot\chi_i=\frac{\partial\chi_i}{\partial t}$ and
$\chi^{\prime}_i=\frac{\partial\chi_i}{\partial x}$.)  Notice that the
classical Hamiltonian coincides with the Hamiltonian obtained \cite{Co} via
Abelian bosonization.  The boundary values of the $\chi_i$ at spatial
infinity are well-known to be related to the electric charge Q (or
equivalently, the "fermion number" B) of the model.  This may be seen by
substituting (4.1) into (2.8) to give
\begin{equation}
B=\frac{1}{e}Q=-\frac{i}{e}\int_{-\infty}^{+\infty} dx
~J^{EM}_2(x,0)=\frac{1}{\sqrt{\pi}}\int_{-\infty}^{+\infty}dx
\sum_i
\chi'_i=\frac{1}{\sqrt{\pi}}\sum_i[\chi_i(\infty,0)-\chi_i(-\infty,0)].
\end{equation}
The equation of motion to be satisfied by the classical ansatz is 
\begin{equation}
\ddot\chi_i-\chi^{''}_i+\frac{e^2}{\pi}(\sum_k\chi_k-\frac{\theta}{2\sqrt{\pi}})+2\sqrt{\pi}m^2_i sin(2\sqrt{\pi}\chi_i)=0.
\end{equation}
For definiteness in what follows we shall specialize to the parity-conserving theory by setting $\theta=0$ and also to the case of two flavors with equal masses.  If we set $\chi_1= -\chi_2\equiv\chi$ the two equations collapse to the sine-Gordon equation
: 
\begin{equation}
\ddot\chi-\chi^{''}+2\sqrt{\pi}m^2 sin(2\sqrt{\pi}\chi)=0.
\end{equation}
Both time-independent and time-dependent classical solutions are important.

{\it i.  Time-independent solution.}  We set  $\dot\chi_i=0$ and choose the
boundary conditions
\begin{equation}
\chi_1(-\infty)=-\chi_2(-\infty)=0,
\qquad \chi_1(\infty)=-\chi_2(\infty)=\sqrt{\pi}.
\end{equation}
Equation (4.3) shows that setting $\chi_1=-\chi_2$ gives zero electric
charge for the solutions.  Then, integrating (4.5) yields the well-known
static sine-Gordon soliton
\begin{equation}
\chi_1(x)=-\chi_2(x)=\frac{2}{\sqrt{\pi}}tan^{-1}exp[2\sqrt{\pi}mx+c'],
\end{equation} 
where $c'$ is an arbitrary constant specifying the soliton location.  The
classical energy is obtained by substituting (4.7) into (4.2):
\begin{equation}
E_{class}=\int_{-\infty}^{+\infty}dx
\ {\cal H}_{class}=\frac{8m}{\sqrt{\pi}} \approx 4.51m.
\end{equation} 
It is amusing that the numerical value of $E_{class}$ is of the same order 
as the lowest-lying $I^{PG}=1^{-+}$ meson mass, $m_\pi$ found in (3.4). 

{\it ii.  Time-dependent solutions.}  There is a well-known \cite{Raj,CoEr} 
family of time-dependent bound solutions of the sine-Gordon equation, 
referred to as "breathers."  They are physically interpreted as a 
bound soliton-anti-soliton pair \footnote{The anti-soliton is obtained by 
giving the right-hand side of (4.7) a negative sign.}.  In our problem, these 
solutions read: 
\begin{equation}
\chi_1(x,t)=-\chi_2(x,t)=\frac{2}{\sqrt{\pi}}
tan^{-1}\left[\frac{a(t)}{cosh(bx)}\right],\quad a(t)=\eta sin\ \omega t,
\quad b=\eta\omega,
\quad \eta=\sqrt{\frac{4\pi m^2}{\omega^2}-1},
\end{equation}
and are characterized by an angular frequency $\omega<2\sqrt{\pi}m$. The
parameter $m$ is the mass which appears in the bosonized Lagrangian.  We obtain
the classical breather energy by substituting (4.9) into (4.2):
\begin{equation}
E_{breather}(\omega)=\int_{-\infty}^{+\infty}dx
\ {\cal H}[\chi_i(x,0)]
=\frac{8\omega}{\pi}\sqrt{\frac{4\pi m^2}{\omega^2}-1}.
\end{equation} 
Both the time-independent classical solution (4.7) as well as the
 time-dependent classical solution (4.9) obey det $U_{c}=1$.  Physically,
 this corresponds to the specialization to the states of the system whose
 masses remain finite as the electric charge $e$ goes to infinity.  As
 discussed in section 3, this means the neglect of the $\eta$ type field
 which can be formally isolated by the decomposition $U=\tilde U
 exp(i\sqrt{\frac{4\pi}{N_f}}\eta)$.  $\tilde U$ satisfies det $\tilde 
U=1$ and describes the light degrees of freedom. 

\section{Semi-classical quantization.}  

In this section we review the semi-classical quantization of the static 
soliton solution in a slightly different way from \cite{Ge} but
with essentially  equivalent results.  We make the ansatz for the matrix
field U \cite{ANW},
\begin{equation}
U(x, t) = A(t)U_c(x, t) A^\dag,
\end{equation}
where A(t) is, in general, an $N_f\times N_f$ special unitary matrix and the
classical solution $U_c$ is allowed, for later purposes, to also depend on
time.  We now substitute this into the bosonized action (2.10).  The first
integral yields
\begin{equation}
...+ \frac{1}{4\pi}\int\ d^2 x\ Tr(U_cA^\dag\dot AU^{\dag}_cA^\dag\dot A-
(A^\dag\dot A)^2+[U^{\dag}_c,\dot U_c]A^\dag\dot A),
\end{equation}
where the three dots stand for the A-independent piece.  Notice that when an
Abelian ansatz like (4.1) is taken, the last term in (5.2) vanishes so no
dependence on $\dot U_c$ remains in the non-classical piece of the
Lagrangian.  For the three-dimensional integral in (2.10) we get
\begin{equation}
\Gamma_{WZW}[U]=\Gamma_{WZW}[A(t)U_c(x,t)A^\dag(t)]=\Gamma_{WZW}[U_c]
+\frac{1}{4\pi}\int d^2 x\ Tr(A^\dag \dot A(U^{\dag}_c
U^{\prime}_c+U^{\prime}_c U^{\dag}_c)).
\end{equation}
The collective variable to be quantized which appears in (5.2) and (5.3) is
clearly $A^\dag(t)\dot A(t)$.  This is an angular-velocity type quantity
which, in the two flavor case of present interest may be decomposed as
\begin{equation}
A^\dag \dot A= \frac{i}{2}\bbox{ \Omega} (t)\bbox{\cdot  \tau},
\end{equation}
where the $\bbox{\tau}$ are the Pauli matrices.  The ansatz in this case
reads
\begin{equation}
U(x,t)=A(t)
\left(\begin{array}{cc}
e^{i2\sqrt{\pi}\chi_1}& 0\\
 0&e^{-i2\sqrt{\pi}\chi_1}
\end{array}\right)
A^\dag(t),
\end{equation}
with $\chi_1$ given in (4.7).  Using (5.2), (5.3) and (5.4) then gives the
collective Lagrangian 
\begin{equation}
L_{coll}=-\frac{8m}{\sqrt{\pi}}+\frac{1}{2}{\cal
I}(\Omega^2_1+\Omega^2_2)-\frac{1}{\sqrt{\pi}}[\chi_1(\infty)
-\chi_1(-\infty)]\Omega_3,
\end{equation}
wherein,
\begin{equation}
{\cal I}= \frac{1}{2\pi}\int_{-\infty}^{\infty} dx\ sin^2[2\sqrt{\pi}
\chi_1(x)]= \frac{2}{3\pi^{3/2}m}.
\end{equation}
The first term in (5.5) represents the classical soliton mass in (4.8).  The
second term comes from (5.2).  The third term comes from (5.3);  using the
boundary condition (4.6) we can see that the coefficient of $\Omega_3$ in
(5.6) is simply -1.  Finally, the quantity in (5.7) will be seen to represent
a ``moment of inertia'' for rotations in isospin space.  It determines the
excitation spectrum and its explicit evaluation is discussed in Appendix A.

The next step is to quantize (5.6).  The canonical momenta (for an implicit
parameterization of the matrix A) may be taken as
\begin{equation}
J_k=\frac{\partial L_{coll}}{\partial\Omega_k}=
\left\{\begin{array}{l}
{\cal I} \Omega_k \qquad k=1,2\\
-1\qquad k=3.
\end{array}\right.
\end{equation}
These yield true dynamical momenta only for $k=1, 2$, but amount to a 
constraint
for $k=3$.  This is analogous to the quantization of the SU(3) Skyrme model
\cite{Gua}.  For quantization we may introduce an operator $J_3$ which,
together with $J_1$ and $J_2$, satisfies the SU(2) algebra
$[J_i,J_j]=i\epsilon_{ijk} J_k$.  However, we must restrict the allowed states to
those obeying
\begin{equation}
J_3|allowed>=-|allowed>.
\end{equation}
The collective Hamiltonian is
\begin{equation}
H_{coll}=-L_{coll}+\frac{\partial L_{coll}}
{\partial\Omega_i}\Omega_i\\
\quad =\frac{8m}{\sqrt{\pi}}+\frac{1}{2{\cal I}}{\bf J}^2-\frac{1}{2{\cal I}}
(J_3)^2.
\end{equation}
After introducing the SU(2) adjoint representation matrix
\begin{equation}
D_{ij}(A)=\frac{1}{2}Tr(\tau_i A\tau_j A^\dag),
\end{equation}
we define $I_i=-D_{ij}(A)J_j$, which can be shown to satisfy ${\bf
I}^2={\bf J}^2$, as well as $[I_i,I_j]=i\epsilon_{ijk}I_k$.  Then, finally,
acting on allowed states, the collective Hamiltonian may be put in the form
\begin{equation}
H_{coll}=\frac{8m}{\sqrt{\pi}}+\frac{1}{2{\cal I}}{\bf I}^2-\frac{1}
{2{\cal I}},
\end{equation}
which describes a rigid rotator in isospin space.  A basis for the space of
states on which the operators in this model act consists of the SU(2)
representation matrices $D^{(I)*}_{m,-m^\prime}(A)$ for isotopic spin $I$.  The
isotopic spin operator {\bf I} acts on the left index of this matrix while
the operator {\bf J} acts on the right index.  The constraint (5.9) requires
that allowed wavefunctions be of the form $D^{(I)*}_{m,1}(A)$ for a meson with
$I_3=m$.  Since a state with $I_3=1$ is evidently required for an allowed
representation we learn that the possible excited states associated with the
classical solution (4.7) are mesons with isospin $I=1, 2, 3,...$ .  In the
fermionic picture these correspond to multi-fermion-anti-fermion states for
$I>1$. In the bosonic picture, the $I>1$ states are bound states of the
fundamental meson in (3.1).

Using the result for the moment of inertia (5.7) in (5.12) finally yields for
the mass of the isospin=I meson,
\begin{equation}
\frac{m(I)}{m}\approx 4.514+ 4.176[I(I+1)-1],\quad I\ge 1.
\end{equation}
The meson with $I=1$ has the mass, $m(1)=8.690m$.  We interpret this meson as
corresponding to the fundamental one in the bosonized Lagrangian.  The mass
obtained by directly reading the coefficient of the quadratic term in (3.3)
is $3.545m$; the different value obtained is interpreted as arising from the
different method of approximation being employed.  Since the fundamental
meson has parity=-1, we expect the parity of meson $I$ to be $(-1)^I$ in the
picture where the meson $I$ is a bound state of $I$ fundamental ones.  Noticing
that $m(2)>2m(1)$, we see that the decay $2\to 1+1$ is energetically allowed.
It is also easy to see that the decay $I\to (I-1)+1$ is energetically
allowed.  Hence, in the present approximation, only the $I=1$ meson is expected
to be stable.

\section{Quantized breather modes and their excitations.}
The quantization of the classical breather solutions in (4.9) is more
involved than the quantization of the static soliton in (4.7).  Whereas the
latter has the fixed mass (4.8), the classical breathers exist for a
continuous family of energies as seen in (4.10).  It is necessary to find the
discrete quantum ``orbits'' by a semi-classical technique like the old Bohr-Sommerfeld method.  Afterwards, one can get excited isotopic spin states by quantization of the variable A(t) in (5.1).  The general picture is very similar to the ``bound state''
 approach to the strange baryons \cite{CK} in the Skyrme model.

A quick way to find the Bohr-Sommerfeld energies was discussed in
\cite{CoEr}.  Since the energy difference between two neighboring
semi-classical (large quantum number n) levels is the classical angular
frequency of periodic motion $\omega$, the number of levels in energy
interval dE will be $dn=\frac{dE}{\omega (E)}$.  Using $\omega (E)$ from
(4.10) and integrating to find n yields
\begin{equation}
E_n=2M\ sin(\frac{n\pi}{8}),
\end{equation}
where $n$ is an integer and $M=\frac{8m}{\sqrt{\pi}}$ is the soliton mass given
in (4.8).  The corresponding angular frequencies are given by
\begin{equation}
\omega_n=(2\sqrt{\pi})m\ cos(\frac{n\pi}{8}).
\end{equation}
Inspection of (6.1) shows that the discrete energies are
$E_1\approx 0.765M,E_2\approx 1.414M$ and $E_3\approx 1.848M$.  The value $E_4=2M$ corresponds to zero angular frequency.
Remember that the breathers are classical solutions of the sine-Gordon
equation (4.5).  In that context, a simple physical interpretation was given
in \cite{Raj_note}.  Expanding the argument of the sine in (6.1) yields
\begin{equation}
E_n\approx n(2m\sqrt{\pi}) + ...\ .
\end{equation}
Now $(2m\sqrt{\pi})$ is recognized from (3.4) as $m_\pi$, the mass of the
fundamental meson degree of freedom of the bosonized theory in the
approximation where meson-meson interactions are neglected.  Thus it is
natural to interpret the $E_n$ solution as a bound state of n fundamental
mesons.  Then the breather solution $E_1$ would be, in fact, a third
alternative description of the fundamental meson.  The breather solution
$E_2$ corresponds to a $\pi\pi$ bound state, etc.  This is the
interpretation adopted by Coleman \cite{Co} in the Abelian quantization case,
and is the one we shall adopt.  Note that the Hamiltonian for our classical
ansatz (4.2) agrees with the Hamiltonian for Coleman's Abelian bosonization
so the classical solution should be the same.  On the other hand, in
ref. \cite{Ge} the breather solution $E_1$ was identified with the $\pi\pi$
bound state.

It should be remarked that a more accurate (argued to be exact) quantization
of the breathers was introduced by Dashen, Hasslacher and Neveu (DHN)
\cite{DHN} and used in \cite{Co}.  It requires the simple modification of
(6.1) to
\begin{equation}
E_n(DHN)=2M\ sin(\frac{n\pi}{6}),
\end{equation}
where $M$ is now the soliton mass with the inclusion of quantum corrections.  
In
this case the only discrete levels are $E_1=M$ and $E_2\approx 1.732M$.  Thus
the level $E_3$ found in the Bohr-Sommerfeld approximation is apparently
spurious.

Now let us consider the semi-classical treatment of the isospin excitations
around (separately) the Bohr-Sommerfeld bound state levels $E_1$ and $E_2$.
We again substitute (5.1), but this time with $U_c$ given by the breather
solution [(4.9) plus (5.5)], into the action.  The work of section 5, in
which $U_c$ was also allowed to depend on time, shows that the analog of the
collective Lagrangian (5.6) becomes
\begin{equation}
L^{(n)}_{coll}=+L^{(n)}_{class}+\frac{1}{2}{\cal I}_n(t)
(\Omega^2_1+\Omega^2_2).
\end{equation}
Here, $L^{(n)}_{class}$ ($n=1, 2$ for the present case) is the classical
Lagrangian whose Legendre transform yields the levels $E_n$ in (6.1).  Note
that the analog of the last term in (5.6) doesn't appear since (4.9) shows
that $\chi_1(x,t=\pm \infty)=0$.  The remaining new feature is that the moment
of inertia depends on time in a complicated way:
\begin{equation}
{\cal I}(t)=\frac{1}{2\pi}\int_{-\infty}^{+\infty}dx\ sin^2\{4\ tan^{-1}[
\frac{\eta\ sin\ \omega t}{cosh(\eta\omega x)}]\},
\end{equation}
where $\eta$ is given in (4.9).  In Appendix A we show that this may be
integrated analytically to yield
\begin{equation}
{\cal I}(t)=\frac{4}{3\pi b}\frac{1}{(a^2+1)^{7/2}}
[-(6a^5-6a^3+3a)ln(\sqrt{a^2+1}-a)+(2a^6-4a^4+9a^2)\sqrt{a^2+1}],
\end{equation}
where $a(t)$ and $b$ are given in (4.9). (Note that the right-hand side actually is an even function of $a$.) 
Plots of ${\cal I}(t)$ for two
particular choices of parameters are shown in Figs. 1 and 2.  As expected, the 
plot looks roughly like a rectified sine curve for the $E_1$ soliton and flattens out into a double square wave as the energy increases.

We will treat (6.5) in an approximate way based on two assumptions.  First,
since the underlying classical motion is periodic, it seems natural to
replace ${\cal I}_n(t)$ by its average over a period $2\pi /\omega_n$:
\begin{equation}
\overline{{\cal I}}_n=\frac{\omega_n}{2\pi}\int_0^{2\pi /\omega_n} dt 
\ {\cal I}(sin\ \omega_n t, b).
\end{equation}
This integral is calculated numerically for the appropriate values of
$\omega_n$.  Secondly, in order to get the correct value isotopic spin=1 for
the fundamental meson it is necessary to consider the collective quantization
component of the isotopic spin ${\bf I}^{coll}$ as an addition to the
isotopic spin of the bound state solution itself ${\bf I}^{bs}$ ($I=1$, $I_3$=0
for the fundamental meson according to \cite{Co}):
\begin{equation}
{\bf I}={\bf I}^{bs}+{\bf I}^{coll}.
\end{equation}

Following section 5 we then obtain the collective Hamiltonian from (6.5) as
\begin{equation}
H^{(n)}_{coll}=E_n+\frac{1}{2\overline{{\cal I}_n} }
({\bf I}^{coll})^2.
\end{equation}
In this case, unlike (5.12), there is no additional restriction on the
allowed values of $({\bf I}^{coll})^2$;  the eigenvalue $({\bf I}^{coll})^2=0$ is now acceptable.

From (6.9), using (4.9) and (6.2), it may be seen that $\overline{ {\cal
I}_n}$ scales as $1/\omega_n$.  Then (6.10) yields a tower of energy levels
for each BS quantized frequency, $\omega_n$:
\begin{equation}
H^{(n)}_{coll}=m[\frac{16}{\sqrt{\pi}}\ sin(\frac{n\pi}{8})
+\frac{\sqrt{\pi}}{\omega_n \overline{ {\cal I}_n}} I_{coll}(I_{coll}+1)\ 
cos(\frac{n\pi}{8})],
\end{equation}
where, from the numerical integration of (6.8),
\begin{equation}
\omega_n\overline{ {\cal I}_n}=\left\{
\begin{array}{lc}
0.742 \qquad n=1\\
0.658 \qquad n=2\\
0.328 \qquad n=3.
\end{array}\right.
\end{equation}
(For comparison, using the DHN frequencies would have given $\omega_1
\overline{{\cal I}_1}=0.796$ and $\omega_2 \overline{{\cal I}_2}=0.426$.)  Let us now examine this spectrum.  First
consider the $n=1$ tower.  The first state has $I^{coll}=0$ and mass $3.457m$.
Using the assumption (6.9) and Coleman's identification $I^{bs}(1)=1$ we get
$I=1$ for this state which is therefore presumed to be yet a third
approximation to the lowest-lying meson state of the model ($I^{PG}=1^{-+}$).
The second level on the $n=1$ tower has $I^{coll}=1$ and mass $7.841m$.  A state
on this level is clearly massive enough to decay to two fundamental mesons and
would then be unstable.  Similarly it is easy to check that all higher states
are heavier than the sum of the preceding level mass and the fundamental
meson mass.  Now consider the $n=2$ tower.  The first state has $I^{coll}=0$
and mass$=6.382m$. Again using (6.9) and Coleman's identification $I^{bs}(2)=0$
we identify this as the $I^{PG}=0^{++}$ meson.  Its mass, according to this
BS quantization is 1.846 times that of the fundamental meson.  (If we had
used the DHN quantization it would be $\sqrt{3}$ times as massive).  The next
level on the $n=2$ tower has a mass$=10.190m$.  It is clearly heavy enough to
decay into $1^{-+}+0^{++}$.  Similarly all higher states of the $n=2$ tower are
massive enough to decay into the preceding one +$1^{-+}$.  The $n=3$ tower will
be considered spurious.

Thus, it seems the present interpretation and approximation in the treatment
of the non-Abelian bosonization can lead to the same stable particle spectrum
as the presumed exact spectrum obtained by Coleman\cite{Co} in the Abelian
bosonization approach using the results of the DHN analysis\cite{DHN}.  In particular, the
numerical values of the averaged moments of inertia obtained are consistent
with the instability of the higher levels on the towers in the
non-Abelian approach.  It would be interesting, however, to introduce
additional ``microscopic'' coordinates associated with the soliton and
anti-soliton components of the breather in order to verify the assumption
(6.9) and to determine all the quantum numbers of the allowed states on the
higher levels.

\section{Additional discussion and an effective Lagrangian}

We have given a different treatment of the non-Abelian bosonized version of
multiflavor $QED_2$ from that presented in \cite{Ge}.  The new features
included are:
i)  starting by ``gauging'' the bosonized free theory (section 2),
ii) using the manifestly symmetric form of the bosonized Lagrangian to
emphasize the analogy to particle physics (section 3),
iii) treating the collective quantization around the static soliton in closer
analogy to the Skyrme model discussions in four-dimensional theories (section
5), and
iv)  a more detailed discussion  of the collective quantization around
the breather solutions (section 6).

It seems worthwhile to summarize the masses of the stable mesons obtained in
the different approximations to the bosonized theory.  Reading the mass term
from the perturbative Lagrangian (3.3) (that is, neglecting interactions)
yields $m(1^{-+})=3.545m$.  Calculating the mass as the first level of the
collective Hamiltonian (5.12) built around the static soliton solution yields
$m(1^{-+})=8.690m$.  Finally, the approximate treatment of the $n=1$ breather
Bohr-Sommerfeld level in (6.11) yields $m(1^{-+})=3.457m$ while the $n=2$ level
yields $m(0^{++})\approx \sqrt{3}m(1^{-+})$.  It is claimed \cite{Co} that the ratio
$m(0^{++})=\sqrt{3}m(1^{-+})$ is exact, while \cite{CoEr} there is no special
reason for different approximation methods to yield especially close results.

Compared to the treatment of the model by Abelian bosonization, the
non-Abelian bosonization is advantageous in getting a general understanding
of the model as illustrated in the treatment of section 3.  On the other
hand, it seems fair to say that treatment of the solitons is more complicated
in the non-Abelian approach.  This is because the extra symmetry of the
Lagrangian introduces extra zero modes, which require collective
quantization in the non-Abelian case.  The virtue of treating solitons in the
non-Abelian approach is that the work may be used to illuminate some aspects
of four-dimensional Skyrme model calculations.  One example of interest is the
study of unequal mass corrections for soliton bound states;  this would be
relevant in the bound state picture of strange and heavy baryons \cite{CK}.
Another example concerns the possible relevance to the Skyrmion treatment of
nucleon-anti-nucleon annihilation \cite{Hol}.

Both the abelian and non-abelian bosonizations yield exact representations of
the fermionic theory.  The non-abelian version has the advantage that the
``charged pions'' ($\pi^\pm$ in the notation of section 3) are present in the
Lagrangian to give manifest isospin invariance (in the two-flavor case).
Now, both versions have the undesirable feature that the other stable
particle in the theory - the $I^{PG}=0^{++}$ particle which we now denote as
$\sigma$ - does ${\it not}$ appear in the Lagrangian.  In fact, it arises in a
rather arcane manner.  This raises the question of whether it is possible to
find a different Lagrangian which also includes the $\sigma$.  This should
not necessarily be an exact representation of the theory but it should be a
good approximation in the ``low-energy'' region.  We shall now see that such
a Lagrangian can be found and furthermore gives a physical motivation for the
basic mass relation
\begin{equation}
m(\sigma)=\sqrt{3}m(\pi).
\end{equation}

We search for an effective Lagrangian which contains the low-lying particles
and respects the underlying symmetries of the exact theory.  It is desired to
model the strong coupling regime of (1.1) taking, for simplicity here, the
two-flavor case.  The underlying symmetry of the massless theory is
manifestly $U(2)_L\times U(2)_R$ which is, however, intrinsically broken to
$SU(2)_L\times SU(2)_R\times U(1)_V$ by quantum corrections (the usual
$U(1)_A$ anomaly).  A common mass term for both flavors will further break
the symmetry to $SU(2)_V$.  Since we are in the strong coupling regime the
$\eta$ particle $(I^{PG}=0^{--})$ is essentially decoupled from the
low-energy theory, as discussed in section 3.  Taking those facts into
account it is clear that the {\it linear} SU(2) sigma model \cite{GL} is a
good candidate to describe low-energy two-flavor $QED_2$.   In this model the
field multiplet contains just the ${\boldmath \pi}$ and $\sigma$ fields as
desired.  We write the Lagrangian density as
\begin{equation}
{\cal L} = -\frac{1}{2}(\partial_\mu \bbox{\pi})^2 - \frac{1}{2}
(\partial_\mu \sigma)^2 - V,
\end{equation}
\begin{equation}
V=A(\sigma^2 + \bbox{\pi}^2 - \lambda)^2 - B\sigma,
\end{equation}
where $A>0$, $B$ and $\lambda$ are three real constants.  The $B\sigma$ term
manifestly breaks the $SU(2)_L\times SU(2)_R$ symmetry down to $SU(2)_V$ and
represents the effect of the fermion mass terms.  We shall assume that this
model is valid for $B=0$ as well as for small $B\ne 0$.  We will work at tree
level here.

To treat this model it is necessary to impose the extremum condition
\begin{equation}
\langle \frac{\partial V}{\partial \sigma}\rangle = 4A\langle \sigma \rangle 
(\langle \sigma \rangle^2 -\lambda) - B=0,
\end{equation}
where $\langle \pi_i \rangle=0$ was taken to agree with parity or isospin
invariance.  We must also demand stability:
\begin{equation}
m^2_\sigma \equiv \langle \frac{\partial^2 V}{\partial \sigma^2}\rangle
=  4A(3\langle \sigma \rangle^2-\lambda) \ge 0,\
m^2_\pi \equiv \langle \frac{\partial^2 V}{\partial \pi_3 \partial \pi_3}
\rangle = 4A(\langle \sigma \rangle^2 - \lambda) \ge 0.
\end{equation}

Now we can see what is special about $m(\sigma)=\sqrt{3}m(\pi)$;
substituting this relation into (7.5) yields the result $\lambda=0$.
Consider the zero mass case $B=0$.  Eq. (7.4) shows there are two possible
solutions \footnote{Note that the usual spontaneous breakdown situation in
the four-dimensional  sigma model corresponds to $\lambda > 0$ and $\langle
\sigma \rangle = \lambda$.  The $\langle \sigma \rangle = 0$ solution is seen
from (7.5) to be unstable.} for the ``condensate'' $\langle \sigma \rangle$:
$\langle \sigma \rangle = 0$ and $\langle \sigma \rangle = \lambda.$  If $\lambda$ vanishes we guarantee that $\langle \sigma \rangle =0$.  This
is, in fact, what is required by the Mermin-Wagner theorem \cite{MW,Shif} which
forbids, for spacetime dimension $\le 2$, a non-zero condensate which
spontaneously breaks a symmetry.  Such an object would signify here the
spontaneous breakdown of chiral SU(2) to $SU(2)_V$ which is not allowed in
two dimensions. (In the one-flavor case the $U(1)_A$ is already
explicitly broken, so these considerations are not relevant.) The situation
is very different from the usual four-dimensional $\sigma$ model in which a
condensate exists for $B=0$ and is maintained as a small non-zero $B$ is turned
on.  The particular mass relation (7.1) is seen to unambiguously force the
unusual two-dimensional behavior.

When $B$ is turned on, the condensate is determined from (7.4) as $\langle
\sigma \rangle=(\frac{B}{4A})^{1/3}$.  Furthermore $m^2_\pi=m^2_\sigma /3=
4A\langle \sigma \rangle^2$.  The present formulation has the nice feature
that it enables the simple calculation of meson scattering amplitudes which
are expected to be accurate in the very low energy region.  For this purpose
we introduce the shifted field
\begin{equation}
\tilde \sigma =\sigma - \langle \sigma \rangle,
\end{equation}
and rewrite the Lagrangian (7.2) as
\begin{equation}
{\cal L} = -\frac{1}{2}[(\partial_\mu \bbox{\pi})^2+(\partial_\mu \tilde
\sigma)^2+ m^2_\pi(\bbox{\pi}^2+3\tilde \sigma^2)]-g_3(\tilde \sigma
\bbox{\pi}\cdot \bbox{\pi}+\tilde \sigma^3)-g_4[(\bbox{\pi}\cdot \bbox{\pi})^2+\tilde \sigma^4+2\bbox{\pi}^2\tilde \sigma^2],
\end{equation}
where $g_3=4A\langle \sigma \rangle$ and $g_4=A$.  Using this Lagrangian we
may compute the tree-level scattering amplitude for
$\pi_i(p_1)+\pi_j(p_2)\to \pi_k(p^\prime_1)+\pi_l(p^\prime_2)$ as
\begin{equation}
A(s,t,u)\delta_{ij}\delta_{kl} +
A(t,s,u)\delta_{ik}\delta_{jl}+A(u,t,s)\delta_{il}\delta_{jk},
\end{equation}
where $s=-(p_1+p_2)^2, t=-(p_1-p^{\prime}_1)^2$, and $u=-(p_1-p^{\prime}_2)^2$,
with
\begin{equation}
A(s,t,u)=-\frac{m^2_\pi}{\langle \sigma
\rangle^2}(\frac{2s-5m^2_\pi}{s-3m^2_\pi}).
\end{equation}

The characteristic feature of this amplitude is the sigma pole below the
threshold at $s_{th}=4m^2_\pi$.  This is in marked contrast to the
four-dimensional case where the sigma mass is not restricted and in fact, the
fairly accurate ``current algebra theorem'' \cite{Wein} is
obtained by taking $m_\sigma \to \infty$.  Because the sigma pole lies so
close to the threshold in the present case, we may reasonably expect it to
dominate the low-energy amplitude.  Loop corrections should become necessary
as one goes away from the threshold region. The accuracy of the model itself away from the threshold region requires more investigation.  Further work beyond these
encouraging initial results, on the low-energy effective Lagrangian approach
to multiflavor $QED_2$ will be reported elsewhere.

\centerline{\bf Acknowledgements}

We would like to thank Francesco Sannino and Herbert Weigel for helpful discussions.  This work was supported in part by the U.S. DOE Contract No. DE-FG-02-ER40231.

\section*{Appendix A}
\setcounter{equation}{0}
\renewcommand{\theequation}{A.\arabic{equation}}

Here we evaluate the moment of inertia integrals.  For the static soliton case
in (5.6),
\begin{equation}
{\cal I}= \frac{1}{2\pi}\int_{-\infty}^{\infty}\ dx\ sin^2 4\theta (x)=
\frac{8}{\pi} \int_{-\infty}^{\infty}\ dx\ (sin^2\theta - 5sin^4\theta +
8sin^6 \theta - 4sin^8 \theta ),
\end{equation}
with $sin^2\theta=1/(1+e^{-4\sqrt{\pi}mx})$, we make the substitution
$y=e^{4\sqrt{\pi}mx}$ to obtain,
\begin{equation}
{\cal I}= \frac{2}{\pi^{3/2}m}\int_0^{\infty}\ dy\ \frac{(1-y)^2}{(1+y)^4}
= \frac{2}{3\pi^{3/2}m}.
\end{equation}

Next consider the time-dependent case in (6.6).  A similar substitution to
the one made above yields
\begin{equation}
{\cal I}= \frac{16}{\pi b}[a^2 I_1-5a^4 I_2+8a^6 I_3-4a^8 I_4],
\end{equation}
where,
\begin{equation}
I_k(a)=\int_1^{\infty}\ \frac{dx}{(x^2+a)^k\sqrt{x^2-1}}.
\end{equation}
From the indefinite integral
\begin{equation}
\int\ \frac{dx}{(x^2+a^2)\sqrt{x^2-1}}=\frac{-1}{a\sqrt{a^2+1}}
ln\left[\frac{x\sqrt{a^2+1}-a\sqrt{x^2-1}}{\sqrt{x^2+a^2}}\right],
\end{equation}
we find
\begin{equation}
I_1(a)= \frac{-ln(\sqrt{a^2+1}-a)}{a\sqrt{a^2+1}}.
\end{equation}
The other $I_k$'s may be obtained by differentiating $I_1(a)$ as
\begin{equation}
I_2(a)=-\frac{I^{\prime}_1}{2a},\quad
I_3(a)=-\frac{1}{8a^3}I^{\prime}_1+\frac{1}{8a^2}I^{\prime\prime}_1,\quad
I_4(a)=-\frac{1}{16a^5}I^{\prime}_1+\frac{1}{16a^4}I^{\prime\prime}_1
-\frac{1}{48a^3}I^{\prime\prime\prime}_1,
\end{equation}
where a prime indicates differentiation with respect to $a$.  Putting
everything together gives (6.7).

\centerline{\bf Figure Captions}

Fig. 1.  Plot of ${\cal I}_1(t)$ for $m=1$.

Fig. 2.  Plot of ${\cal I}_2(t)$ for $m=1$.

%\begin{center}
%\epsfysize=160pt
%\epsfxsize=10pt
%\ \epsfbox{Fig_1_exp.eps}
%\begin{itemize}
%\item[Fig. 1]
%{${\cal I}$ vs. t}
%\end{itemize}
%\end{center}
 
\end{document}